\documentclass[11pt,a4paper]{article}
\begin{document}

\title{ Is the Schwarzschild black hole really stable? \footnote{E-mail of Tian:
xinda-2002@126.com, hua2007@126.com, tgh-2000@263.net}}
\author{Guihua Tian$^{1,2}$, Shikun Wang$^{2}$,\ Zhao Zheng$^{3}$\\
1.School of Science, Beijing University \\
of Posts And Telecommunications. Beijing100876, China.\\2.Academy
of Mathematics and Systems Science,\\ Chinese Academy of
Sciences,(CAS) Beijing, 100080, P.R. China,\\ 3.Department of
Physics, Beijing Normal University, Beijing100875, China.}
\date{April 18, 2005}
\maketitle

\begin{abstract}
The stability of the Schwarzschild black hole is studied. Regge
and Wheeler treated the problem first at 1957 and obtained the
dynamical equations for the small perturbation. There are two
kinds of perturbations: odd one and even one. Using the
Painlev\'{e} coordinate, we reconsider the odd perturbation and
find that: the \textbf{\emph{white-hole-connected universe}}
($r>2m$, see text) is unstable. Because the odd perturbation may
be regarded as the angular perturbation, therefore, the physical
mean to it may be that the \textbf{\emph{white-hole-connected
universe}}  is unstable with respect to the rotating perturbation.

\textbf{PACC:0420-q}
\end{abstract}

Studies concerning the stability of the Schwarzschild black hole
may step back to the work of the Regge and Wheeler, who first
divided the perturbation into odd and even ones \cite{rw}. Later,
it is found that odd one is really the angular perturbation to the
metric, while even one corresponds to the radial perturbation to
the metric \cite{chan}.

Vishveshwara made the study further by transforming the
perturbation quantities to the Kruskal reference frame, and tried
to find the real divergence at $ r=2m $ from the spurious one
caused by the improper choice of coordinate due to the
Schwarzschild metric's ill-defined-ness at $ r=2m $ \cite{vish}.
Later, Price also studied the problem carefully \cite{pric} and
Wald studied from the mathematical background \cite{wald}.

In reference \cite{stew}, Stewart applied the Liapounoff theorem
to define dynamical stability of a black-hole. First, according
Stewart, the normal mode of the perturbation fields to the
Schwarzschild black-hole is the perturbation fields $\Psi$ with
time-dependence of $e^{-ik t}$ which are bounded at the boundaries
of the event horizon $r=2m$  and the infinity $r\rightarrow \infty
$. The range of permitted frequency is defined as the spectrum $S$
of the Schwarzschild black-hole. Then, for the Schwarzschild
black-hole, it could be obtained by  the Liapounoff theorem
that\cite{stew}:

(1)if $\exists k \in S$ with $\Im k >0$, the Schwarzschild
black-hole is dynamically unstable,

(2)if $\Im k <0$ for $\forall k \in S$ , and the normal modes are
complete, then, the Schwarzschild black-hole is dynamically
stable,

(3)if $\Im k \leq 0$ for $\forall k \in S$  , and there is at
least one real frequency $k \in S$, the linearized approach could
not decide the stability of the Schwarzschild black-hole.

Vishveshwara proved the normal mode of the Schwarzschild
black-hole could be real. Therefore, strictly speaking, the
stability of the Schwarzschild black-hole is unsolved according to
this criterion of Stewart's definition.

Here, we reconsider the  stability problem of the Schwarzschild
black hole using the Painlev\'{e} coordinate metric(see
following). The conclusions are:  for the
\textbf{\emph{white-hole-connected universe}} of the Schwarzschild
space-time(see following), there exists frequencies with $\Im k
>0$. So, the \textbf{\emph{white-hole-connected universe}} of the Schwarzschild
space-time is unstable.

Recently, in studying the Hawking radiation as tunnelling
\cite{Pari}, the Schwarzschild coordinates is replaced by the
Painlev\'{e} coordinates, which were discovered independent by
Painlev\'{e} in 1921\cite{pain} and Kraus, Wilczck in 1994
\cite{Krau}, \cite{Krau1}. The Painlev\'{e} metric is stationary
and regular at the horizon \cite{Krau}, \cite{Krau1}. This good
quality makes it more suitable for studying the stability of the
Schwarzschild black hole.

In the following, we first introduce the Painlev\'{e} coordinate
metric for the Schwarzschild black hole, then using the
Vishveshwara's result, we transform the odd perturbation
quantities to the Painlev\'{e} coordinate system,  and study the
problem.

The Schwarzschild metric is
\begin{equation}
ds^{2}=-(1-\frac{2m}{r})dt_{s}^{2}+(1-\frac{2m}{r})^{-1}dr^{2}+r^{2}
d \Omega ^{2}.\label{orimetric}
\end{equation}
By
\begin{equation}
t_{p}=t_{s}-\left[2\sqrt{2mr}+2m \ln
\frac{\sqrt{r}-\sqrt{2m}}{\sqrt{r}+\sqrt{2m}}\right],\label{paintran1}
\end{equation}
we obtain the Painlev\'{e} metric of the black hole,
\begin{eqnarray}
ds^{2} &=& -\left(1-\frac{2m}{r}\right)dt_p^{2}
-2\sqrt{\frac{2m}{r}}drdt_p+dr^{2}+r^{2}d\Omega ^{2} \nonumber\\
&=&
-dt_p^{2}+\left(dr-\sqrt{\frac{2m}{r}}dt_p\right)^{2}+r^{2}d\Omega
^{2}.\label{painwhitemetric}
\end{eqnarray}
The Painlev\'{e} metric (\ref{painwhitemetric}) is well-behaved at
the horizon. It is obvious that the radial null geodesic are
\begin{equation}
\frac{dr}{dt_p}=\pm 1+ \sqrt{\frac{2m}{r}}
\end{equation}
respectively. The upper and lower classes correspond to the
outgoing and ingoing null geodesics. In the usual Penrose diagram,
part $I$ corresponds to our universe($ r>2m $), parts $II$ and
$II'$ are the black-hole($ r<2m $) and white-hole($ r<2m $)
respectively, and part $I'$ is another universe($ r>2m $) not
communicating with our universe.

Clearly, the metric (\ref{painwhitemetric}) represents the region
$I$ and $II'$ in the Penrose diagram. We define the universe $I$
connected with $II'$ by the metric (\ref{painwhitemetric}) as the
\textbf{\emph{white-hole-connected universe}}.

If we transform the metric (\ref{orimetric}) by
\begin{equation}
t_{p}=t_{s}+\left[2\sqrt{2mr}+2m \ln
\frac{\sqrt{r}-\sqrt{2m}}{\sqrt{r}+\sqrt{2m}}\right],\label{paintran2}
\end{equation}
we obtain another Painlev\'{e} metric of the black hole:
\begin{eqnarray}
ds^{2} = -\left(1-\frac{2m}{r}\right)dt_p^{2}
+2\sqrt{\frac{2m}{r}}drdt_p+dr^{2}+r^{2}d\Omega
^{2}\label{painblackmetric}
\end{eqnarray}
The Painlev\'{e} metric (\ref{painblackmetric}) is also regular at
the horizon. The radial outgoing and ingoing null geodesics are
\begin{eqnarray}
\frac{dr}{dt_p} = \pm 1-\sqrt{\frac{2m}{r}}
\end{eqnarray}
respectively. Similarly, the metric (\ref{painblackmetric})
represents the region $I$ ($ r>2m $,our universe) and $II$($ r<2m
$, the black hole) in the Penrose diagram. We define the universe
$I$ connected with $II$ by the metric (\ref{painblackmetric}) as
the \textbf{\emph{black-hole-connected universe}}.

Here, we briefly explain the perturbation  of the Schwarzschild
black hole.

Suppose the background metric is $g_{\mu \nu}$, while the
perturbation in it is $h_{\mu \nu}$, the contract Ricci tensors
$R_{\mu \nu}$, $R_{\mu \nu}+\delta R_{\mu \nu} $ correspond the
metrics $g_{\mu \nu}$, $g_{\mu \nu}+h_{\mu \nu} $ respectively.
The non-linear perturbation equation is
\begin{equation}
\delta R_{\mu \nu}=0,\label{rab0}
\end{equation}
while the linear part  with respect to $h_{\mu \nu}$ in the
equation (\ref{rab0}) is  called the perturbation field equation.
After the consideration of the gauge freedom, the odd perturbation
is
$$
h_{\mu \nu}=\left|
\begin{array}{cccc}
0&                0&   0&  h_{0}(r)\\
0&                0&   0&  h_{1}(r)\\
0&                0&   0&  0       \\
h_{0}(r)&  h_{1}(r)&   0&  0
\end{array}
\right|   e^{-ikt_s}\left[\sin \theta \frac{\partial}{\partial
\theta}\right]P_{l}\left(\cos \theta\right),
$$
and the even one is
$$
h_{\mu \nu}=\left|
\begin{array}{cccc}
H_{0}(1-\frac{2m}{r})&    H_{1}&   0&  0\\
H_{1}&   H_{2}(1-\frac{2m}{r})^{-1}&   0&  0\\
0&                0&   r^{2}K&  0       \\
0&  0&   0&  r^{2}K \sin^{2}\theta
\end{array}
\right|   e^{-ikt_s}P_{l}\left(\cos \theta\right).
$$

Here we mainly discuss the odd perturbation.

In the odd perturbation, the linear equations of (\ref{rab0}) are
combined into one single equation, that is, the Regge-Wheeler
equation:
\begin{equation}
\frac{d^{2}Q}{dr^{*2}}+\left[k^{2}-V\right]Q=0,\label{ReggeWheeler}
\end{equation}
where the effective potential $V$ and the tortoise coordinate are
\begin{equation}
V=\left(1-\frac{2m}{r}\right)\left[\frac{l\left(l+1\right)}{r^{2}}-\frac{6m}{r^{3}}\right],
\end{equation}
\begin{equation}
r^{*}=r+2m\ln \left(\frac{r}{2m}-1\right)\label{tortoise}
\end{equation}
respectively. The perturbation field $h_{0}(r)$ and $h_{1}(r)$ are
connected with $Q$ by
\begin{equation}
h_{0}(r)=\frac{i}{k} \frac{d}{dr^{*}}\left(rQ\right)=\frac{i}{k}
\left[\left(1-\frac{2m}{r}\right)Q+r\frac{dQ}{dr^{*}}\right]\label{h0}
\end{equation}
and
\begin{equation}
h_{1}(r)=r\left(1-\frac{2m}{r}\right)^{-1}Q.\label{h1}
\end{equation}

The Schwarzschild black hole is unstable if initially
well-behaved, the perturbation fields are growing in time. In the
odd perturbation case, that is, $h_{0}(r)$ and $h_{1}(r)$ are
well-behaved initially in the range $(2m, \infty)$ or
$(-\infty,+\infty)$ by the tortoise coordinate $r^*$, and at least
one of them run into infinity when time increases to infinity.
Just as stated before, if the imaginary part $\Im k$ of the
frequency $k$ is greater than zero, then the black hole is
unstable with respect to the perturbation. Vishveshwara showed
$\Re k=0,\Im k>0$ is impossible in reference \cite{vish}. Here, we
will prove that $\Re k=0,\Im k>0$ is possible for the
\textbf{\emph{white-hole-connected universe}}.

As stated before, the Schwarzschild metric is singular at $r=2m$,
so, we could not discuss the problem under this metric. The metric
(\ref{painwhitemetric}) is regular at the horizon, so we study the
perturbation under this metric. Just as done in the reference
(\cite{vish}), we  get the perturbation field quantities in the
Schwarzschild metric coordinates for simplification, then we
obtain the corresponding quantities in the metric
(\ref{painwhitemetric}) of Painlev\'{e} coordinates by
transformation and study the stable problem.

By equation (\ref{paintran1}), it is easy to get the perturbation
fields $[h^{p}_{ij}]$ in the metric (\ref{painwhitemetric}) of
Painlev\'{e} coordinates:
\begin{equation}
h_{03}^{p}=h_{03}^{s}
\end{equation}
\begin{equation}
h_{13}^{p}=h_{13}^{s}+\sqrt{\frac{2m}{r}}\left(1-\frac{2m}{r}\right)^{-1}h_{03}^{s}
\end{equation}
where
\begin{equation}
h_{03}^{s}=h_{0}e^{-ikt_s}=\frac{i}{k}
\left[(1-\frac{2m}{r})Q+r\frac{dQ}{dr^{*}}\right]e^{-ikt_s}
\end{equation}
\begin{equation}
h_{13}^{s}=h_{1}e^{-ikt_s}=r\left(1-\frac{2m}{r}\right)^{-1}Qe^{-ikt_s}
\end{equation}
(see equations (\ref{h0}) and (\ref{h1})), therefore,
\begin{equation}
h_{03}^{p}=\frac{i}{k}\left[\left(1-\frac{2m}{r}\right)Q+r\frac{dQ}{dr^{*}}\right]e^{-ikt_s}\label{h0white}
\end{equation}
\begin{equation}
h_{13}^{p}=\left[\frac{i}{k}\sqrt{\frac{2m}{r}}Q+r\left(1-\frac{2m}{r}\right)^{-1}
\left(\frac{i}{k}\sqrt{\frac{2m}{r}}\frac{dQ}{dr^{*}}+Q\right)\right]e^{-ikt_s}\label{h1white}
\end{equation}

Now, we solve the differential equation (\ref{ReggeWheeler})
\begin{equation}
\frac{d^{2}Q}{dr^{*2}}+\left[k^{2}-V\right]Q=0
\end{equation}

As proved in reference \cite{vish}, when $k=ik_{2}$, $k_{2}>0$,
the asymptotic solutions (\ref{ReggeWheeler}) as $r^{*}\rightarrow
\infty$ are $ \tilde{A}e^{\pm k_{2}r^{*}}$, and the well-behaved
one is $\tilde{A}e^{-k_{2}r^{*}}$, that is,
\begin{equation}
Q_{\infty}=\tilde{A} e^{-k_{2}r^{*}}
\end{equation}
and from the equation (\ref{ReggeWheeler}), the solution
$\tilde{A}e^{-k_{2}r^{*}}$ cannot become $Ae^{+k_{2}r^{*}}$ as
$r^{*}\rightarrow -\infty $. Therefore the asymptotic solution to
$r^{*}\rightarrow -\infty $ is \cite{vish}
\begin{equation}
Q_{2m}=Ae^{-k_{2}r^{*}}\label{infty}.
\end{equation}

It is easy to see that $Q_{2m}$ is singular at the horizon $ r=2m
$, or, $r^{*}\rightarrow -\infty $. Nevertheless, it may be caused
by the ill-behaved-ness of the Schwarzschild metric
(\ref{orimetric}) at the horizon $ r=2m $. We transform the
perturbation fields to  the Painlev\'{e} coordinate system, and
will prove the subsequent perturbation fields are initially
well-defined at $r=2m$.

Substituting equation (\ref{infty}) into (\ref{h0white}) and
(\ref{h1white}), and using the transformation equation
(\ref{paintran1}), then,  at $ r=2m$, that is, $ r^{*}\rightarrow
-\infty $, it is easy to get
\begin{eqnarray}
&&h^{p}_{03}(t_{p},r) = \frac{1}{k_{2}}\left[(1-\frac{2m}{r})-k_{2}r\right]Ae^{k_{2}(t_s-r^{*})}   \\
 &=& \frac{1}{k_{2}}\left[(1-\frac{2m}{r})-k_{2}r\right]Ae^{k_{2}t_{p}}
 e^{k_{2}\left[-r+2\sqrt{2mr}-4m\ln\left(1+\sqrt{\frac
r{2m}}\right)\right]}\label{h03-2}
\end{eqnarray}
and
\begin{eqnarray}
h^{p}_{13}(t_{p},r) &=&
\frac{1}{k_{2}}\sqrt{\frac{2m}{r}}Ae^{k_{2}(t_s-r^{*})}
+r\left(1-\frac{2m}{r}\right)^{-1}\left[\frac{1}{k_{2}}\sqrt{\frac{2m}{r}}(-k_{2})+1\right]
Ae^{k_{2}(t_s-r^{*})}   \\
 &=& \left[\frac{1}{k_{2}}\sqrt{\frac{2m}{r}}+r(1+\sqrt{\frac{2m}{r}})^{-1}\right]
 Ae^{k_{2}t_{p}}
 e^{k_{2}\left[-r+2\sqrt{2mr}-4m\ln\left(1+\sqrt{\frac
r{2m}}\right)\right]}.\label{h13-2}
\end{eqnarray}
From equation (\ref{h03-2}) and (\ref{h13-2}), it is easy to see
that $h^{p}_{13}$, $h^{p}_{03}$ are regular at $r=2m$ initially(at
$t_{p}=0$), so, $h^{p}_{13}(0,r)$ and $h^{p}_{03}(0,r)$ are
regular in $(2m,\infty)$, or $(-\infty ,+\infty)$ by $r^{*}$.
Therefore, the \textbf{\emph{white-hole-connected universe}} of
the Schwarzschild space-time is unstable with respect to this kind
perturbation.

This completes our proof.

We restate our conclusions again: the
\textbf{\emph{white-hole-connected universe}} of the Schwarzschild
space-time is unstable with respect to the angular perturbations.
Besides, our method could be used in study of the stable problem
of Reissner-Nordstr\"{o}m black-hole, similar results could be
obtained.

\section*{Acknowledgments}

We are supported in part by the National Science Foundation of
China under Grant No.10475013, No.10375087, No.10373003,
No.10375008, NKDRTC(2004CB318000) and the post-doctor foundation
of China.


\begin{thebibliography}{99}
\bibitem{rw} T. Regge, J. A. Wheeler, Phys. Rev.{\em 108},  1063 ,(1970).
\bibitem{chan} S. Chandrasekhar,  The mathematical theroy of black hole. Oxford: Oxford University
Press,(1983).
\bibitem{vish}  C. V. Vishveshwara, Phys. Rev.{\em D1},  2870 ,(1957).
\bibitem{pric}  R. Press, Phys. Rev.{\em D5},  2419 ,(1972).
\bibitem{wald}  R. M. Wald, J. Math. Phys.{\em 20}, 1056 ,(1979).
\bibitem{stew} J. M. Stewart, Proc. R. Soc. {\em A344},   65 ,(1975).
\bibitem{Pari} M. K.  Parikh, F. Wilczek, Phys. Rev. Lett.{\em 85},  5042 ,(2000).
\bibitem{pain}  P. Painlev\'{e}, C. R. Hebd, Seances Acad. Sci.{\em 173}, 677 ,(1921).
\bibitem{Krau} P. Kraus, F. Wilczek, Mod. Phys. Lett.{\em A9},  3713 ,(1994).
\bibitem{Krau1} P. Kraus, F. Wilczek, Nucl. Phys.{\em B433},   403 ,(1996).
\end{thebibliography}
\end{document}